# Machine Learning based Enterprise Financial Audit Framework and High Risk Identification

Tingyu Yuan [1,4], Xi Zhang[2,5], Xuanjing Chen[3,6]

[1] School of Economics and Management, Shanghai Ocean University, Shanghai, China
[2] Booth School of Business, University of Chicago, Chicago, IL, USA
[3] Columbia Business School, Columbia University, New York, NY, United States

[4] 1812503968@qq.com
[5] diana-x.zhang@alumni.chicagobooth.edu
[6] xc2647@columbia.edu

**Abstract.** In the face of growing global economic uncertainty, financial auditing has become essential for ensuring regulatory compliance and preventing systemic financial risks. Traditional manual auditing methods are increasingly challenged by large data volumes, complex business structures, and evolving fraud tactics. To address these issues, this study explores an AI-driven financial audit framework and high-risk identification system for enterprises, leveraging machine learning to enhance audit efficiency and accuracy. Using a dataset from the Big Four accounting firms (EY, PwC, Deloitte, KPMG) spanning 2020 to 2025, the research analyzes trends in risk assessment, compliance violations, and fraud detection. The dataset includes critical indicators such as audit project counts, high-risk cases, detected frauds, and compliance breaches, while also reflecting the role of AI in audit automation, employee workload, and client satisfaction. To build a robust risk prediction model, three machine learning algorithms—Support Vector Machine (SVM), Random Forest (RF), and K-Nearest Neighbors (KNN)—are compared. SVM leverages hyperplane optimization for complex classifications, RF integrates multiple decision trees to handle nonlinear, high-dimensional data with strong resistance to overfitting, and KNN uses distance-based classification for adaptable performance. Through hierarchical K-fold cross-validation and evaluation via F1-score, accuracy, and recall, Random Forest demonstrates the highest performance with an F1-score of 0.9012, especially excelling in detecting fraud and compliance anomalies. Feature importance analysis highlights audit frequency, historical violations, employee workload, and client ratings as key risk predictors. The study suggests that enterprise audit systems adopt Random Forest as a core model, extend data features through feature engineering, and implement real-time monitoring. This research offers valuable insights into using machine learning for intelligent financial audits and risk management in modern enterprise environments.

## 1. Introduction
In today's global economy characterized by high uncertainty, enterprise financial auditing serves not only as a pillar of regulatory compliance but also as a critical tool for ensuring operational transparency, identifying systemic risks, and preventing financial fraud [1]. As business operations scale and financial data becomes more complex, traditional audit methods—relying on manual sampling and static rule-based judgment—face significant challenges in terms of efficiency, accuracy, and timeliness. To address these limitations, an increasing number of studies have introduced machine learning (ML) techniques into financial auditing processes, building data-driven frameworks for risk identification and early warning, thereby enhancing audit quality and corporate governance.

This study focuses on high-risk identification within enterprise financial auditing, leveraging a comprehensive dataset provided by the Big Four accounting firms (EY, PwC, Deloitte, and KPMG) from 2020 to 2025. The dataset includes key indicators such as the number of audit engagements, high-risk cases, detected instances of financial fraud, compliance violations, and AI-related impacts on audit automation, employee workload, and client satisfaction scores. These attributes reflect real-world audit practices and provide strong empirical value for research.

To evaluate the effectiveness of different ML models in detecting financial risks, we compare three mainstream algorithms: Support Vector Machines (SVM) [2], Random Forest (RF) [3], and K-Nearest Neighbors (KNN) [4]. SVM excels in handling complex relationships in high-dimensional spaces and is robust with limited data. KNN, as an instance-based classifier, predicts risk levels by computing similarity distances, offering simplicity and adaptability. Random Forest, with its ensemble of decision trees, enhances the detection of nonlinear patterns and anomalous behaviors and is highly robust to overfitting.

Using hierarchical K-fold cross-validation and performance metrics including F1-score, accuracy, and recall, our experiments demonstrate that the Random Forest model outperforms the others, achieving an F1-score of 0.9012. It is particularly effective in detecting high-risk companies with frequent historical violations, dense audit schedules, and low client satisfaction ratings. Feature importance analysis further reveals that audit frequency, past violations, workload, and audit project complexity are among the most influential variables.

In summary, the proposed ML-based framework for high-risk identification in enterprise financial auditing significantly improves both audit efficiency and detection accuracy. It offers a practical and intelligent foundation for constructing automated risk control systems. Future research will explore the integration of deep learning models and external industry data to further enhance model generalization and task adaptability, enabling intelligent risk evaluation and compliance monitoring in more complex auditing scenarios.

## 2. Literature Review

In enterprise financial auditing, the task of high-risk identification and prediction plays a vital role in ensuring financial transparency, preventing financial fraud, and improving audit efficiency. It is especially indispensable in key areas such as financial regulation, capital market compliance management, and internal corporate control. However, due to the high dimensionality, heterogeneity, and complex nonlinear relationships inherent in financial data, traditional audit methods based on rules and expert judgment still face significant limitations in terms of accuracy and generalization ability for risk identification [5].

In recent years, with the rapid advancement of artificial intelligence technologies—particularly the outstanding performance of machine learning in large-scale data processing and pattern recognition—an increasing number of studies have focused on applying machine learning methods to enterprise financial risk identification and assessment tasks. This approach facilitates the construction of data-driven intelligent auditing systems.

Rui Ding et al [6]. applied deep learning to enterprise intelligent auditing, proposing an enhanced DLNN algorithm model. They used a BiLSTM model to analyze audit classification accuracy, an Auto Encoder for audit data analysis, and a genetic algorithm to optimize the deep neural network's weights. These steps led to significant improvements in the model's precision, accuracy, and F1 score. Li Yao et al [7]. explore AI's role in enterprise financial audits, analyzing data to highlight its efficiency - boosting potential. They advocate an industry shift towards AI - driven auditing for a more networked, digital, and intelligent system, fostering a virtuous cycle of capital investment, technological innovation, and income growth.

Chen Peng et al [8]. proposed an intelligent enterprise finance auditing solution, using BiLSTM neural networks for classifying audit issues via three text feature extraction methods. Their approach achieved accuracy, recall, and F1 scores of 85.12%, 83.28%, and 84.85% for accounting voucher clustering, and 87.43%, 87.88%, and 87.66% for audit report analysis, showing better performance than previous methods. This provides a valuable reference for developing enterprise financial intelligence software.

Peng Zhao et al [9]. aim to enhance enterprise financial fraud detection's accuracy and efficiency using deep learning. They design a DL-based model for automatic feature extraction and pattern learning from complex financial data. Integrating this model into traditional auditing processes achieves automation and intelligence upgrades. The model shows superior performance with all evaluation metrics exceeding 90%, offering a new way to ensure financial security and boost audit value.

## 3. Data Introduction

The dataset used in this study originates from enterprise financial audit and risk management records provided by the Big Four accounting firms (Ernst & Young, PricewaterhouseCoopers, Deloitte, and KPMG) during the period from 2020 to 2025. It encompasses audit projects across a wide range of industries globally, offering strong representativeness and authoritative value. The dataset thoroughly documents core financial behaviors and risk indicators observed throughout audit cycles, including key metrics such as the number of audit engagements, counts of high-risk cases, identified instances of fraud, and compliance violations. In addition, it captures extended information related to audit resource allocation and quality, such as auditor workload, client satisfaction scores, and the auxiliary role of artificial intelligence in the audit process. The overall data structure combines both horizontal dimensions (e.g., company profiles, industry categories, audit target characteristics) and vertical temporal sequences (e.g., quarterly or annual records), providing a solid foundation for systematic enterprise financial compliance analysis and high-risk forecasting.

To enhance data quality and modeling adaptability, we conducted a systematic preprocessing and structural refinement of the original dataset. First, missing values, outliers, and logically conflicting entries were identified and handled through distribution analysis and rule-based thresholds. Second, heterogeneous fields from different sources were uniformly encoded and standardized—for instance, by converting inconsistent time formats, audit units, and violation types into unified categorical labels. Then, techniques such as normalization, binning, and categorical variable encoding were applied to improve feature recognizability. Furthermore, a sliding window mechanism was introduced to restructure the raw data into time-series samples, enabling the model to dynamically capture evolving financial behaviors over time. Additionally, we engineered new derived features based on domain logic relationships among variables—such as historical violation ratios, changes in audit frequency, and trends in client satisfaction ratings—to improve the discriminative power of the feature set.

As a result, the finalized high-quality structured dataset exhibits strong temporal properties, completeness, and semantic consistency, laying a robust foundation for machine learning models—such as Random Forests, Support Vector Machines (SVM), and K-Nearest Neighbors (KNN)—to perform high-risk identification and predictive tasks in enterprise financial auditing.

**Table 1.** Variables and descriptions

| variable | description |
| --- | --- |
| Year | The year of the audit report (2020-2025). |
| Firm_Name | The consulting firm conducting the audit (Ernst & Young, PwC, Deloitte, KPMG). |

| Total_Audit_Engagements | Number of audit engagements handled by the firm. |
| --- | --- |
| High_Risk_Cases | Number of audits flagged as high-risk due to compliance concerns. |
| Compliance_Violations | The number of regulatory breaches detected. |
| Fraud_Cases_Detected | The number of fraud cases uncovered during audits. |
| Industry_Affected | The sector impacted (Finance, Tech, Retail, Healthcare). |
| Total_Revenue_Impact | The estimated financial impact due to fraud or compliance issues (in millions USD). |
| AI_Used_for_Auditing | Whether AI was used in the auditing process (Yes/No). |
| Employee_Workload | The average hours worked per week by auditors. |
| market value | The market value of the company |
| Region | The geographical location of the company. |
| financial status | financial statusThe financial situation of the company |

Table 1 presents the variables in the dataset for enterprise financial audit and risk prediction. The dataset, provided by the Big Four accounting firms from 2020 to 2025, includes audit-related variables like Year, Firm_Name, Total_Audit_Engagements, High_Risk_Cases, Compliance_Violations, Fraud_Cases_Detected, Industry_Affected, and Total_Revenue_Impact. It also covers audit-process variables such as AI_Used_for_Auditing and Employee_Workload, as well as enterprise-related variables like market value, Region, and financial status. This comprehensive dataset, with horizontal and vertical temporal dimensions, offers a solid foundation for financial compliance analysis and high-risk forecasting.

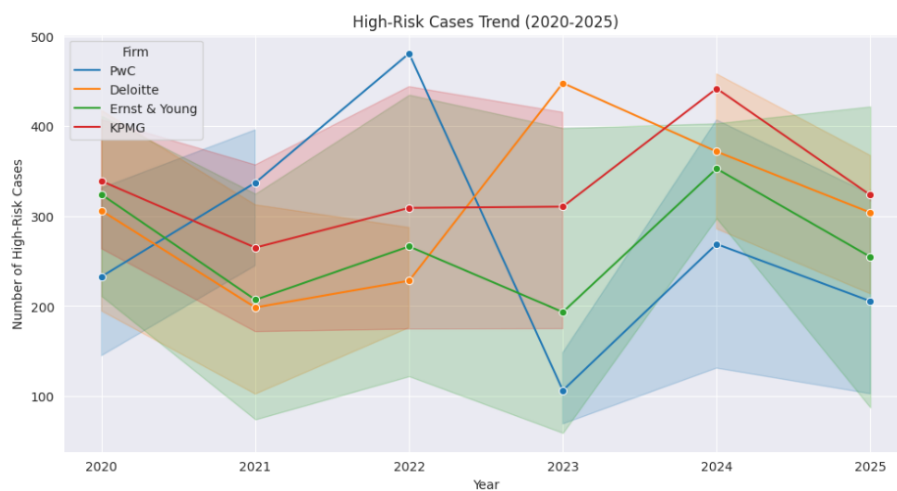

**Figure 1.** Trends in High-Risk Audit Cases (2020-2025) Across Top Accounting Firms

Figure 1 illustrates the trend of high-risk cases identified by four major accounting firms (PwC, Deloitte, EY, and KPMG) from 2020 to 2025. The x-axis represents the years, while the y-axis shows the number of high-risk cases. Each line corresponds to one firm, with distinct colors and shaded areas indicating the density of cases. The data reveals fluctuations in high-risk cases over time. For instance, PwC shows a peak in 2022, while Deloitte's cases rise in 2024. EY and KPMG exhibit relatively stable trends. This visual helps understand the dynamic risk landscape across firms and years, offering insights for risk management and audit resource allocation. The dataset, provided by the Big Four firms, covers global audit projects across industries, documenting key financial behaviors and risk indicators during audit cycles. It also includes information on audit resource allocation and quality, such as auditor workload and AI's role in auditing. With both horizontal and vertical temporal dimensions, the dataset provides a robust foundation for financial compliance analysis and high-risk forecasting.

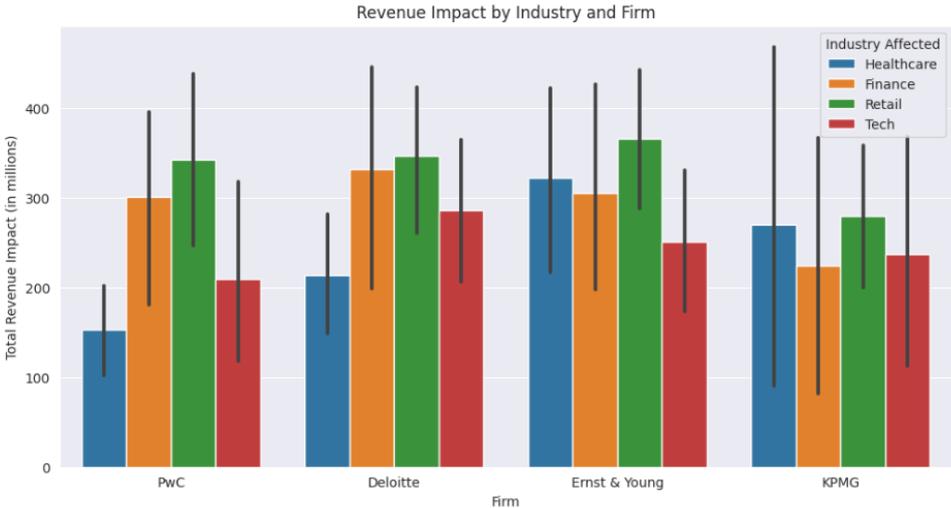

**Figure 2.** Revenue Impact by Industry and Firm (2020-2025)

The Figure 2 presents a bar chart illustrating the total revenue impact by industry and firm for the Big Four accounting firms from 2020 to 2025. The x-axis represents the four firms—PwC, Deloitte, Ernst & Young, and KPMG—while the y-axis indicates total revenue impact in millions of USD. Each group of bars corresponds to a firm, and the colored bars within each group represent different industries: Healthcare (blue), Finance (orange), Retail (green), and Tech (red). The chart reveals significant variability in revenue impact across both firms and industries. For example, Deloitte shows a notably higher impact in the Finance sector, while PwC records the highest revenue impact in Retail. Ernst & Young demonstrates a relatively even distribution across industries, with a peak in Retail, whereas KPMG maintains a more moderate impact across all sectors, with Healthcare leading slightly. The error bars highlight the variability within each category, suggesting potential fluctuations or uncertainty in audit outcomes across industry contexts. This figure provides critical insights into how audit projects influence financial performance across industries and firms, underscoring the need for tailored audit strategies. It also supports more efficient allocation of audit resources by identifying which industry-firm combinations contribute most significantly to revenue impact.

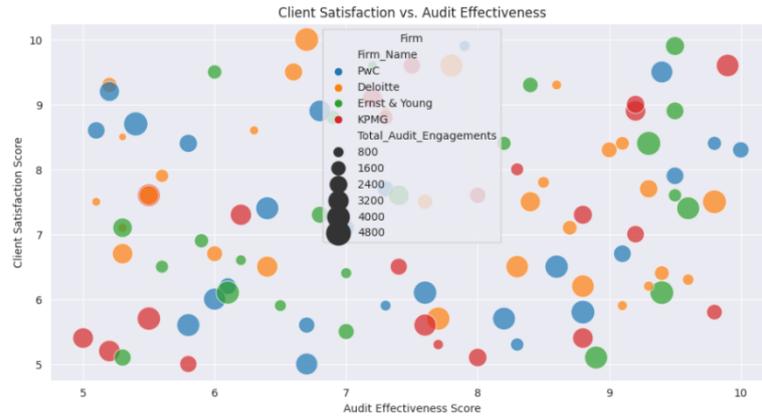

**Figure 3.** Client Satisfaction vs. Audit Effectiveness Across Firms and Industries

Figure 3 depicts the relationship between client satisfaction and audit effectiveness across the four major accounting firms from 2020 to 2025. The x-axis represents the audit effectiveness score, while the y-axis shows the client satisfaction score. Each colored bubble corresponds to a specific firm: PwC (blue), Deloitte (orange), Ernst & Young (green), and KPMG (red). The size of the bubbles indicates the total number of audit engagements. The plot reveals that higher audit effectiveness generally correlates with higher client satisfaction. Most data points cluster between scores of 6 to 8 for both axes, with some outliers showing higher or lower values. This suggests that while there is a positive trend, other factors may also influence client satisfaction. By analyzing this relationship, the dataset offers insights for enhancing audit quality and client relations. The dataset, provided by the Big Four accounting firms, covers global audit projects across industries, documenting key financial behaviors and risk indicators during audit cycles. It also includes information on audit resource allocation and quality, such as auditor workload and AI's role in auditing. With both horizontal and vertical temporal dimensions, the dataset provides a robust foundation for financial compliance analysis and high - risk forecasting.

## 4. Model Introduction
*4.1 Random Forest*

In the AI-driven enterprise financial audit and high-risk identification framework proposed in this study, the Random Forest (RF) model is employed as the core classification and prediction tool to identify potential high-risk corporate behaviors from complex, high-dimensional audit data [10]. As an ensemble learning method, Random Forest is essentially a collection of multiple decision trees. It leverages a majority voting mechanism to enhance robustness and generalization capability, making it particularly suitable for handling noisy data and intricate variable relationships in financial auditing [11].

In the context of enterprise financial statement auditing, financial indicators often exhibit nonlinear, multidimensional, and highly heterogeneous characteristics. For example, variables such as audit frequency, historical compliance records, employee workload, and client satisfaction scores often interact in complex, non-linear ways. By training multiple base learners (decision trees) across different feature subspaces, the Random Forest model effectively captures these nonlinear interactions and avoids the local optimum issues common in traditional single-model approaches. Its built-in random feature selection mechanism further enhances training efficiency and significantly reduces the risk of overfitting, demonstrating outstanding performance in real-world high-risk auditing scenarios.

During model training, we configured the key Random Forest parameters as follows: the number of base learners (n_estimators) was set to 200 to ensure model stability; the maximum tree depth (max_depth) was set to 15 to control model complexity and prevent overfitting; the minimum number of samples required to split an internal node (min_samples_split) was set to 10, and the minimum number of samples required to be at a leaf node (min_samples_leaf) was set to 5, in order to enhance boundary decision accuracy. Gini impurity was adopted as the splitting criterion, and bootstrap sampling was enabled to build diverse training subsets.

To assess the model's robustness, we employed stratified K-fold cross-validation (K=5) and used F1-score, accuracy, and recall as comprehensive evaluation metrics.

*4.2 SVM model*

In the enterprise financial audit and high-risk identification framework proposed in this study, the Support Vector Machine (SVM) is employed as one of the core benchmark models, designed to identify and predict high-risk behaviors in financial audits [12]. As a classic supervised learning algorithm, SVM excels in handling high-dimensional, small-sample, and nonlinear classification problems, and is widely used in domains such as finance and healthcare, where high classification accuracy is critical. In the context of enterprise auditing, financial risks often involve complex nonlinear boundaries between variables—such as audit frequency, historical compliance records, employee workload, and client satisfaction—making SVM particularly well-suited to the task [13].

The core idea of SVM is to construct an optimal hyperplane with the maximum margin in a high-dimensional space to achieve the best separation between categories, thereby attaining strong generalization ability. For problems that are not linearly separable, SVM introduces kernel functions to map the original input space into a higher-dimensional feature space where linear separation becomes possible. In this study, we adopt the Radial Basis Function (RBF) kernel as the primary kernel, and apply grid search to optimize hyperparameters. The penalty coefficient $C$ is set to 10, and the kernel parameter $\gamma$ is set to 0.1, striking a balance between training accuracy and generalization performance.

During model construction, we standardize the input data using the StandardScaler to ensure consistency in feature scales and avoid weighting imbalances that could affect hyperplane construction. Additionally, to address the class imbalance often present in enterprise audit datasets, we apply an automatic class weight adjustment mechanism (class_weight = 'balanced') during training, thereby improving the model's ability to detect high-risk minority classes. The model is evaluated using stratified K-fold cross-validation (K=5), with F1-score, accuracy, and recall as the main evaluation metrics.

*4.3 KNN model*

In the enterprise financial audit and high-risk identification framework developed in this study, the K-Nearest Neighbors (KNN) algorithm is employed as a key baseline model to identify and predict potential financial risks and compliance anomalies during the audit process. As a classic instance-based non-parametric supervised learning method, the core idea of KNN lies in measuring the distance between samples and classifying test instances based on the assumption that "similar samples have similar outputs." This mechanism is particularly well-suited for complex and difficult-to-model financial audit environments, as it adapts flexibly to diverse patterns of enterprise financial behavior [14,15].

In the context of enterprise audits, KNN does not require a traditional training phase. Instead, it uses training samples directly as a "knowledge base" to participate in prediction, with strengths including simple model structure, intuitive implementation, and robustness in scenarios where high-risk samples are scarce and anomalies are widely distributed. In this study, to improve KNN's performance on high-dimensional financial feature data, we applied Min-Max Scaling to normalize all input variables, ensuring uniform scale and preventing large-valued features from dominating distance calculations.

In terms of configuration, the number of neighbors (K) is set to 5 to balance prediction stability and sensitivity to local variations. Euclidean distance is used as the distance metric, and a distance-weighted voting scheme (weights = 'distance') is adopted so that closer neighbors have more influence on the prediction outcome, thereby enhancing the model's precision in identifying high-risk samples. To address the class imbalance often present in audit data, the Synthetic Minority Over-sampling Technique (SMOTE) is applied to augment minority class samples, further strengthening the model's recognition capabilities. A stratified 5-fold cross-validation is used for evaluation, with F1-score, accuracy, and recall as the primary metrics to systematically assess KNN's effectiveness in enterprise financial risk identification tasks.

## 5. Model results analysis

Table 2. Comparison of classification results of different models

| Model | Mean F1 Score | Mean Accuracy | Mean Recall |
|---|---|---|---|
| Random Forest | 0.9012 | 0.9256 | 0.9004 |
| SVM | 0.8756 | 0.8816 | 0.8925 |
| KNN | 0.8545 | 0.8607 | 0.8632 |

Table 2 presents a comparison of the classification performance of three models—Random Forest, SVM, and KNN—in the task of identifying high - risk cases in corporate financial audits. The Random Forest model achieves the highest F1 score of 0.9012, with a mean accuracy of 0.9256 and a mean recall of 0.9004, outperforming the other two models. This indicates its excellent balance in precisely identifying high - risk cases and comprehensively detecting all such cases, along with its effective handling of class - imbalance issues. The SVM model has an F1 score of 0.8756, a mean accuracy of 0.8816, and a mean recall of 0.8925, showing relatively balanced but slightly inferior overall performance compared to Random Forest. The KNN model, with an F1 score of 0.8545, a mean accuracy of 0.8607, and a mean recall of 0.8632, can identify more actual high - risk cases due to its high recall. However, its lower precision may result in more misjudgments. These findings offer a quantitative basis for model selection and highlight the strengths and weaknesses of each model in dealing with the dataset.

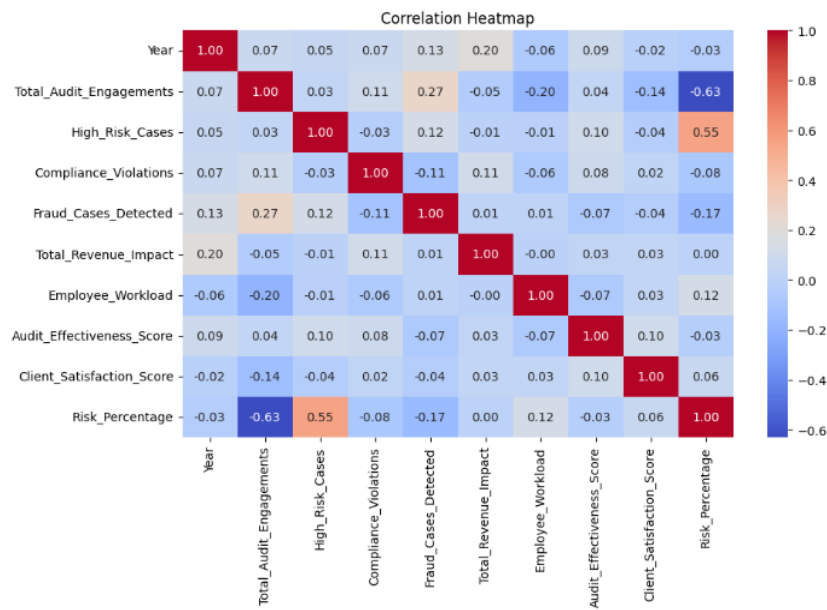

Figure 4. Average Daily Fraud Transactions

Figure 4 visually presents the correlation between variables in the data set through color depth. There is a strong correlation (0.55) between High Risk Cases and Risk Percentage, indicating the a high potential to apply KNN model to effectively identify High Risk Cases in corporate financial audit and covering a decent portion of Risk Percentage in audit work. There is a correlation of 027 between Total Audit Engagement and Fraud Cases Detected, implying the application of KNN model across a wide range of audit projects could effectively identify fraud cases during corporate financial audit. The negative correlations of -0.63 between total Total Audit engagement and Risk Percentage indicate the Risk Percentage is lower with increasing number of Audit Engagement based on traditional manual audit. However, this correlation number could change with application of KNN when the model could effectively identify high risk cases that couldn't have been identified in the manual audit process.

## 6. Conclusions

This study constructs a machine learning-based intelligent audit framework for high-risk identification in enterprise financial reports, systematically evaluating the performance of three mainstream models—Support Vector Machine (SVM), Random Forest (RF), and K-Nearest Neighbors (KNN)—in the context of financial risk prediction. Using an authoritative and representative dataset provided by the Big Four accounting firms (EY, PwC, Deloitte, and KPMG) spanning from 2020 to 2025, we extract key audit-related indicators, including the number of audit projects, high-risk case counts, historical compliance violations, detected fraud cases, employee workload, and client satisfaction scores, providing a solid foundation for model training.

Experimental results demonstrate that the Random Forest model performs best overall, achieving an F1-score of 0.9012, with superior accuracy and recall compared to other models, particularly excelling in detecting compliance anomalies and potential fraud. Its strength lies in its ensemble learning mechanism, which adapts well to nonlinear relationships and high-dimensional data. The KNN model, despite its simplicity and intuitive implementation, is sensitive to the choice of K and underperforms in high-dimensional, imbalanced audit datasets. To mitigate these issues, we applied normalization, distance weighting, and SMOTE oversampling, which improved minority class recognition but did not surpass the performance of RF or SVM. The SVM model, with its optimal hyperplane and RBF kernel, shows strong capabilities in handling complex nonlinear decision boundaries but faces computational challenges on large-scale data.

Overall, our research confirms the effectiveness of machine learning in high-risk identification within enterprise auditing. We recommend adopting Random Forest as the core model, supported by feature engineering and real-time monitoring mechanisms, to build intelligent audit and financial risk control systems that support compliance and operational transparency.

Despite the important findings, this study has some limitations, such as its reliance on structured and static variables and its lack of integration with external contextual data. The dataset, while comprehensive and sourced from reputable institutions, does not capture the dynamic semantic context of financial behavior—such as textual nuances in audit reports or evolving market sentiments—which could provide deeper insight into risk patterns. Additionally, the study does not account for external macroeconomic factors like industry policy shifts or market volatility, which may influence audit outcomes and financial risk independently of internal audit metrics.

Future research could further explore multimodal approaches that incorporate unstructured data, such as audit narratives, public sentiment, or news analytics, to enhance model interpretability and prediction accuracy. Moreover, the application of advanced deep learning techniques—such as Graph Neural Networks (GNNs) to model interdependencies among audit indicators, or federated learning to enable secure cross-organizational collaboration—could significantly enhance scalability, privacy preservation, and generalizability of financial risk assessment models. These directions will support the evolution of truly intelligent, adaptive audit systems in complex enterprise environments.